# Solução da Equação de Poisson-Boltzman Aplicação na Membrana do Neurônio Humano

[1]**Marilia Amável Gomes Soares**
[2]**Frederico Alan de Oliveira Cruz**
[3]**Célia Martins Cortez**

[1]*Escola de Ciências da Saúde, Universidade do Grande Rio*
[2]*Mestrado Profissional em Ensino de Ciências na Educação Básica, Universidade do Grande Rio*
[3]*Departamento de Fisiologia, Universidade do Estado do Rio de Janeiro*

**Resumo:** Com já demonstrado em trabalho anterior as equações que descrevem a dependência espacial do potencial elétrico são determinadas pela solução da equação de Poisson-Boltzmann. Neste trabalho propomos essas soluções para a membrana do neurônio humano, utilizando um modelo simplificado para essa estrutura considerando a distribuição dos eletrólitos em cada lado da membrana, bem como o efeito do glicocálix e da bicamada lipídica. Nós assumimos que em ambos os lados da membrana as cargas estão homogeneamente distribuídas e que o potencial depende apenas das coordenada *z*.

**Abstract:** With already demonstrated in previous work the equations that describe the space dependence of the electric potential are determined by the solution of the equation of Poisson-Boltzmann. In this work we consider these solutions for the membrane of the human neuron, using a model simplified for this structure considering the distribution of electrolytes in each side of the membrane, as well as the effect of glicocálix and the lipidic bilayer. It was assumed that on both sides of the membrane the charges are homogeneously distributed and that the potential depends only on coordinate z.

## 1. Introdução

Em todos os estudos já realizados sobre as concentrações iônicas das células de mamíferos foi observado que existe uma significante diferença entre os meios extra e intracelular. Tal diferença é mantida pela própria célula, pois, muitas vezes, uma substância que tem pouca importância funcional num dos lados da membrana é de grande importância no outro [1]. Um exemplo desse fenômeno ocorre com os íons de sódio ($Na^+$) e potássio ($K^+$), já que este último é mais abundante dentro da célula do que no meio extracelular, sendo o contrário no caso do primeiro.

Devido às diferentes concentrações iônicas se estabelecem fluxos iônicos, naturalmente produzidos pelos princípios da equiosmolaridade e eletroneutralidade, que regem o equilíbrio entre as soluções intra e extracelulares. Dessa forma há fluxo constante de $Na^+$ para o interior da célula e um escoamento constante de $K^+$ para o fluido externo [2]. Entretanto, apesar do movimento desses íons favorecido pelos seus gradientes de concentração através da membrana, esses gradientes são mantidos, graças a um mecanismo que se opõe a este movimento. No caso destes íons, essa diferença de concentração é mantida pela quebra de uma molécula de *ATP* e com a participação de uma proteína $Na^+$-$K^+$-adenosina-trifosfatase, comumente chamada bomba de $Na^+$-$K^+$. Neste processo são "trocados" três $Na^+$ por dois $K^+$, como representado pela cinética abaixo:

$$ATP + 3Na^+_{in} + 2K^+_{out} \leftrightarrow ADP + Pi + 3Na^+_{out} + 2K^+_{in} \qquad (1)$$

onde os índices *in* e *out* representam os íons nos meios intra e extracelular [3], *ATP*, *ADP* e *Pi* são adenosina trifosfato, adenosina difosfato e fosfato inorgânico respectivamente [4].

Além da importância funcional que esses íons possuem em cada lado da membrana, o gradiente existente está envolvido num grande número de processos, incluindo a manutenção do potencial de membrana, sendo a contribuição do $Na^+$, do $K^+$ e do $Cl^-$ ao potencial descrita pela expressão simplificada Goldman-Hodgkin-Katz (*GHK*):

$$\phi_{GHK} = \frac{RT}{F} \ln\left( \frac{P_K[K^+_{out}] + P_{Na}[Na^+_{out}] + P_{cl}[Cl^-_{in}]}{P_K[K^+_{in}] + P_{Na}[Na^+_{in}] + P_{cl}[Cl^-_{out}]} \right) \quad (2)$$

onde *V* representa o potencial em função da concentração iônica dos íons dentro e fora da célula, $P_{Na}$, $P_K$ e $P_{Cl}$, as permeabilidades desses íons em relação a membrana, e *R*, *T* e *F* representam a constante universal dos gases, temperatura (em Kelvin) e a constante de Faraday respectivamente [5].

A grande limitação da equação *GHK* é que ela descreve apenas o potencial transmembranar, mas não o perfil de potencial existente em cada lado. O perfil descreve o comportamento do potencial em todos os pontos adjacentes ao longo de uma direção perpendicular à membrana e a sua determinação pode ser efetuada através de modelos teóricos [6,7,8,9] que levam em consideração as contribuições dos íons em cada lado da membrana e também das cargas fixas nas regiões adjacentes [10,11,12].

Neste trabalho analisamos a distribuição iônica dos íons $Na^+$, $K^+$ e $Cl^-$ a partir da equação que descreve o perfil de potencial da membrana do neurônio, levando em conta a permeabilidade da membrana a esses íons no estado de equilíbrio.

## 2. Modelo adotado

A membrana do neurônio é formada por uma bicamada fosfolipídica, contendo em sua estrutura ainda colesterol e proteínas [13], onde no arranjo dessa bicamada, estão colocados de cada lado da membrana, de forma contínua, uma cadeia de carboidratos [14]. Considerando esses conhecimentos e das estruturas celulares adotaremos para o nosso estudo o modelo mostrado na Figura 1.

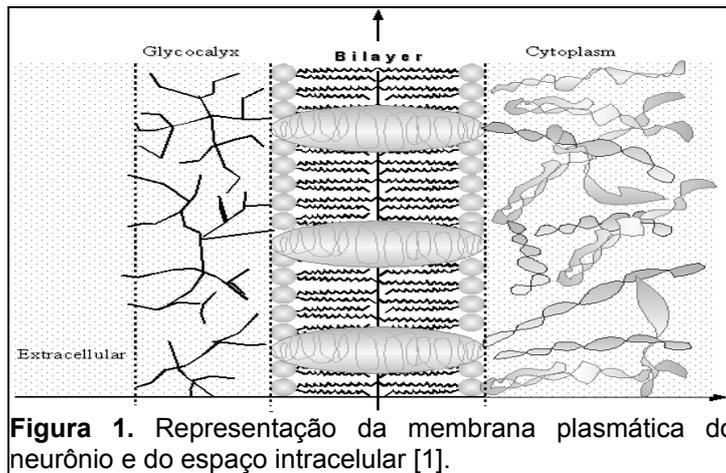

**Figura 1.** Representação da membrana plasmática do neurônio e do espaço intracelular [1].

Esse sistema é formado por quatro regiões, meio extracelular, glicocálix, bicamada lipídica e citoplasma, onde cada uma delas contribui para o potencial em cada ponto da membrana. Supondo que o potencial depende apenas das contribuições em uma determinada direção, o primeiro passo é determinar uma expressão que descreva o potencial em função da distância, para o potencial nas quatro regiões (Figura 2).

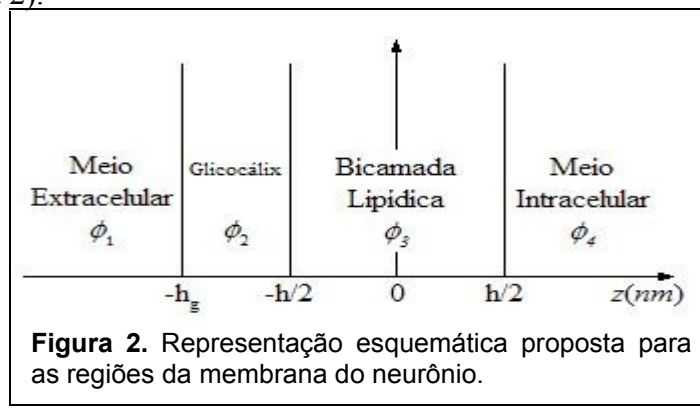

**Figura 2.** Representação esquemática proposta para as regiões da membrana do neurônio.

Como estamos supondo que o sistema possua comportamento eletrostático, a lei de Faraday será reescrita como:
$$\vec{\nabla} \times \vec{E} = 0 \qquad (3)$$
o que nos permite definir que o campo elétrico está relacionado ao potencial escalar da forma:

$$\vec{E} = -\vec{\nabla}\phi \qquad (4)$$

Mas pela lei de Gauss na forma diferencial, temos que:
$$\vec{\nabla} \cdot \vec{D} = \rho \qquad (5)$$
onde, $D$, é o vetor deslocamento elétrico e $\rho$ é a densidade de carga na região considerada. Pela equação constitutiva $D = \varepsilon E$ e pela equação (4), supondo que o meio seja homogêneo e isotrópico, a equação (5) é reescrita na forma:
$$\vec{\nabla}^2 \phi = -\frac{\rho}{\varepsilon} \qquad (6)$$
chamada de equação de Poisson para o potencial escalar [15]. Como nosso problema trata de de quatro regiões com distribuições volumétricas de cargas distintas, podemos escrever a equação (6) na forma:
$$\vec{\nabla}^2 \phi_i = -\frac{\rho_i}{\varepsilon_i} \qquad (7)$$
onde $\rho_i$ representa a densidade de cargas, $\phi_i$ o potencial e $\varepsilon_i$ a permissividade em cada região, onde $i = 1$ representa o meio extracelular, $i = 2$ a região do glicocálix, $i = 3$ a bicamada lipídica e $i = 4$ o meio intracelular.

Num sistema biológico composto por cargas imersas em meio aquoso contendo eletrólitos, a carga total do meio será dada pela soma algébrica das cargas fixas do meio, $\rho_f$, e das cargas dos eletrólitos dissolvidos no solvente, consideradas como cargas móveis, $\rho_m$, determinada pela distribuição de Boltzmann [16]:
$$\rho_{mi} = \sum_j q_{ij} \eta_{ij} \exp\left[-\frac{q_{ij}\phi_i}{k_B T}\right] \qquad (8)$$
onde $q$ é a carga correspondente a cada íon, $\eta$ é a densidade de íons do tipo $j$ por unidade de volume na região considerada, $\phi$ é o potencial eletrostático na região considerada, $k_B$ é a constante de Boltzmann e T a temperatura absoluta.

No caso das regiões 1 e 4 consideramos que a distribuição volumétrica de cargas, $\rho$, é dada apenas pela distribuição de Boltzmann não havendo cargas fixas, fornecendo para essas regiões a expressão:
$$\vec{\nabla}^2 \phi_i = -\frac{1}{\varepsilon_i} \sum_j q_{ij} \eta_{ij} \exp\left[-\frac{q_{ij}\phi_i}{k_B T}\right] \text{ onde } i = 1 \text{ ou } 4 \qquad (9)$$
para a região 2, que refere-se ao glicocálix, supomos haver uma distribuição fixa de cargas além das cargas móveis que eventualmente atravessam estrutura, sendo assim temos que:
$$\vec{\nabla}^2 \phi_2 = -\frac{1}{\varepsilon_2}\left(\rho_f + \sum_j q_{2j}\eta_{2j} \exp\left[-\frac{q_{2j}\phi_2}{k_B T}\right]\right) \qquad (10)$$
Agora, considerando apenas os íons monovalentes e a condição de eletroneutralidade para a distribuição, tal qual como o realizado para a membrana do eritrócito [17], podemos reescrever as expressões (9) e (10) como:
$$\vec{\nabla}^2 \phi_i = -2\frac{q_i \eta_{0i}}{\varepsilon_i} \operatorname{senh}\left[\frac{q_i \phi_i}{k_B T}\right] \text{ onde } i = 1 \text{ ou } 4 \qquad (11)$$
$$\vec{\nabla}^2 \phi_2 = -\frac{1}{\varepsilon_2}\left(\rho_f + 2q_2 \eta_2 \operatorname{senh}\left[\frac{q_2 \phi_2}{k_B T}\right]\right) \qquad (12)$$

que são as equações de Poisson-Boltzmann para as regiões 1, 4 e 2 respectivamente. Finalmente para a região 3, que designa a bicamada lipídica, consideramos que $\rho = 0$, visto que a densidade de cargas elétricas positivas e negativas são iguais, fornecendo assim:

$$\vec{\nabla}^2 \phi_3 = 0 \qquad (13)$$

Ficamos, assim, de posse de todas as equações necessárias para a determinação do perfil de potencial da nossa distribuição de cargas, para a membrana do neurônio.

### 3. Condições de assimetria e contorno

Para determinarmos o potencial em cada ponto do nosso modelo, inicialmente consideramos apenas que o potencial sofrerá variação apenas na direção $z$ e a homogeneidade de cargas nas direções $x$ e $y$ [10]. Sendo assim as expressões do Laplaciano dependem apenas das distribuições em $z$, fornecendo:

$$\vec{\nabla}^2 \phi_i = \frac{d^2 \phi(z)}{dz^2} \qquad (14)$$

Introduzimos também as seguintes condições de normalização baseados em dados da literatura [2] onde $\phi(-\infty) \neq \phi(\infty)$ visto que:

(i) No meio extracelular, em $z \to -\infty$, o potencial será:
$$\phi(-\infty) = 0$$

(ii) e no meio intracelular, em $z \to -\infty$, o potencial será:
$$\phi(\infty) = -70 \text{ mV}$$

esse valores representam, respectivamente, os potencias nos limites interno e externo.

Como existem cargas nas superfícies de separação das regiões, haverá, portanto, descontinuidade na componente normal do campo elétrico, tal que:

Condição de contorno da derivada normal
$$\begin{cases} \varepsilon_2 \phi_2'(-h_g) - \varepsilon_1 \phi_1'(-h_g) = \sigma_g \\ \varepsilon_3 \phi_3'\left(-\frac{h}{2}\right) - \varepsilon_2 \phi_2'\left(-\frac{h}{2}\right) = \sigma_1 \\ \varepsilon_4 \phi_4'\left(\frac{h}{2}\right) - \varepsilon_3 \phi_3'\left(\frac{h}{2}\right) = \sigma_2 \end{cases} \qquad (15)$$

onde $\sigma_i$ é a densidade superficial de carga nas superfícies que separam os meios.

A última das considerações necessárias para a obtenção das expressões está relacionada à descontinuidade sofrida pelo potencial elétrico devida a diferença de permeabilidade das regiões:

Condição de contorno do potencial
$$\begin{cases} \varepsilon_1 \phi_1(-h_g) = \varepsilon_2 \phi_2(-h_g) \\ \varepsilon_2 \phi_2\left(-\frac{h}{2}\right) = \varepsilon_3 \phi_3\left(-\frac{h}{2}\right) \\ \varepsilon_3 \phi_3\left(\frac{h}{2}\right) = \varepsilon_4 \phi_4\left(\frac{h}{2}\right) \end{cases} \qquad (16)$$

### 4. Solução para as fases externa e interna

Para as regiões internas e externas as membrana a expressão a ser resolvida é a eq.(11), considerando a condição de homogeneidade das cargas nas direções $x$ e $y$, tal que:

$$\frac{d^2 \phi_i(z)}{dz^2} = -2 \frac{q_i \eta_i}{\varepsilon_i} \text{senh}\left[\frac{q_i \phi_i(z)}{k_B T}\right] \qquad (17)$$

escrevendo de maneira mais simplificada, chamando:

$$a_i = -2\frac{q_i \eta_i}{\varepsilon_i} \qquad (18)$$

$$b_i = \frac{q_i}{k_B T} \qquad (19)$$

Com o auxílio de (18) e (19), obtemos a eq.(17) na forma:

$$\frac{d^2\phi_i(z)}{dz^2} = a_i \,\text{senh}[b_i\phi_i(z)] \qquad (20)$$

A solução da eq.(20), em primeira ordem, nos fornece a expressão para o campo elétrico nas regiões internas e externas a membrana:

$$E_i(z) = -\frac{d\phi_i(z)}{dz} = -\sqrt{2\frac{a_i}{b_i}\cosh[b_i\phi_i(z)] + \lambda_i} \qquad (21)$$

onde $\lambda_i$ é uma constante a ser determinada pela condições de contorno da respectiva membrana e que possui unidade de medida em volt por metro para que a mesma possa ser consistente com a análise dimensional da eq(21). Introduzindo a relação:

$$\frac{a_i}{b_i} = 2\frac{\eta_i k_B T}{\varepsilon_i} = \xi_i \qquad (22)$$

Voltando a equação (21) devemos lembrar que para regiões superiores a 1,5 nm da membrana o potencial deve ser contante, logo o campo elétrico deve ser nulo. Para a fase 1, meio externo, optamos pela análise do cosseno hiperbólico utilizando uma série de Taylor, em torno do ponto onde $\phi_i = 0$:

$$\cosh[b_1\phi_1(z)] = 1 + \frac{1}{2}b_1^2\phi_1^2 + \ldots \qquad (23)$$

substituindo (23) em (21), teremos:

$$E_1(z) = -\frac{d\phi_1(z)}{dz} = \mp\sqrt{2\frac{a_1}{b_1}\left[1 + \frac{1}{2}b_1^2\phi_1^2(z) + \ldots\right] + \lambda_1} \qquad (24)$$

Para $\phi_1(z) = 0$ devemos ter $E_1(z) = 0$, então obtemos que:

$$\lambda_1 = -2\frac{a_1}{b_1} \qquad (25)$$

fornecendo a relação:

$$E_1(z) = -\frac{d\phi_1(z)}{dz} = \mp\sqrt{a_1 b_1[\phi_1(z)]^2 + \ldots} \qquad (26)$$

Após uma análise numérica dos termos e desprezando o termos de ordem superiores, pois contribuem pouco para os resultados encontrados, $\phi_i^2$ podemos escrever:

$$E_1(z) = -\frac{d\phi_1(z)}{dz} \approx \mp\sqrt{a_1 b_1}\,\phi_1(z) \qquad (27)$$

Resolvendo a equação (27), obtemos a expressão para o potencial elétrico:

$$\phi_1(z) = \phi_{1_{01}} e^{\sqrt{a_1 b_1}\,z} + \phi_{1_{02}} e^{-\sqrt{a_1 b_1}\,z} \qquad (28)$$

Para a fase 4, meio interno, a série de Taylor foi expandida em torno do ponto $\phi_4 = -70$ mV, o que nos fornece:

$$\cosh[b_4\phi_4(z)] = \cosh(-70 b_4) + \text{senh}(-70 b_4)b_4[\phi_4(z) + 70] + \cosh^2(-70 b_4)\frac{b_4^2[\phi_4(z) + 70]^2}{2} + \ldots \qquad (29)$$

Substituindo (29) na expressão (24) obtemos para a região 2:

$$E_4(z) = -\frac{d\phi_4(z)}{dz} = \mp\sqrt{2\frac{a_4}{b_4}\cosh(-70 b_4) + 2a_4\,\text{senh}(-70 b_4)[\phi_4(z) + 70] + \frac{a_4 b_4}{2}\cosh^2(-70 b_4)[\phi_4(z) + 70]^2 + \ldots + \lambda_4} \qquad (30)$$

Para $\phi_4(z) = -70$ devemos ter $E_4(z) = 0$, então obtemos que:

$$\lambda_1 = -2\frac{a_4}{b_4}\cosh(-70b_4) \quad (31)$$

fornecendo a relação:

$$E_4(z) = -\frac{d\phi_4(z)}{dz} = \mp\sqrt{2a_4\operatorname{senh}(-70b_4)[\phi_4(z)+70] + \frac{a_4 b_4}{2}\cosh^2(-70b_4)[\phi_4(z)+70]^2 + \ldots} \quad (32)$$

Fazendo a primeira substituição onde $V(z) = \phi_4(z) + 70$, teremos:

$$\frac{dV(z)}{dz} = \pm\sqrt{2a_4\operatorname{senh}(-70b_4)V(z) + \frac{a_4 b_4}{2}\cosh^2(-70b_4)V(z)^2 + \ldots} \quad (33)$$

Desprezando os termos de ordem superior a 2 pois novamente eles pouco influenciam em nosssos resultados, obtemos:

$$\frac{dV(z)}{dz} \approx \pm\sqrt{2a_4\operatorname{senh}(-70b_4)V(z) + \frac{a_4 b_4}{2}\cosh^2(-70b_4)V(z)^2} \quad (34)$$

Transformando a expressão acima num quadrado perfeito, teremos:

$$\frac{dV(z)}{dz} \approx \pm\sqrt{\left(\sqrt{\frac{a_4 b_4}{2}}\cosh(-70b_4)V(z) + \sqrt{\frac{2a_4}{b_4}}\tanh(-70b_4)\right)^2 - \frac{2a_4}{b_4}\tanh^2(-70b_4)} \quad (35)$$

Colocando o termo que possui a tangente hiperbólica em evidência, teremos:

$$\frac{dV(z)}{dz} \approx \pm\sqrt{\frac{2a_4}{b_4}}\tanh(-70b_4)\sqrt{\left(\frac{b_4}{2}\frac{\cosh(-70b_4)}{\tanh(-70b_4)}V(z) + 1\right)^2 - 1} \quad (36)$$

Fazendo a segunda substituição:

$$l(z) = \frac{b_4}{2}\frac{\cosh(-70b_4)}{\tanh(-70b_4)}V(z) + 1 \quad (37)$$

Usando (37) em (36), obtemos:

$$\frac{dl(z)}{dz} \approx \pm\sqrt{\frac{a_4 b_4}{2}}\cosh(-70b_4)\sqrt{l(z)^2 - 1} \quad (38)$$

Resolvendo (38), chegamos a:

$$l(z) = \cosh\left[\sqrt{\frac{a_4 b_4}{2}}\cosh(-70b_4)(h_\infty - z)\right] \quad (39)$$

Usando (37) em (39) e lembrando que $V(z) = \phi_4(z) + 70$, chegamos finalmente a:

$$\phi_4(z) = -70 + \frac{2\tanh(-70b_4)}{b_4\cosh(-70b_4)}\cosh\left[\sqrt{\frac{a_4 b_4}{2}}\cosh(-70b_4)(h_\infty - z) - 1\right] \quad (40)$$

## 5. Solução para o glicocálix

Para a região do glicocálix a expressão a ser resolvida, eq.(12), será reescrita como:

$$\frac{d^2\phi_2(z)}{dz^2} = -\frac{1}{\varepsilon_2}\left(\rho_f + 2Z_2 e\eta_2\operatorname{senh}\left[\frac{q_2\phi_2(z)}{k_B T}\right]\right) \quad (41)$$

definindo:

$$G = -\frac{\rho_f}{\varepsilon_2} \quad (42)$$

e utilizando as equações (18) e (19), podemos reescrever (27) como:

$$\frac{d^2\phi_2(z)}{dz^2} = G + a_2\operatorname{senh}[b_2\phi_2(z)] \quad (43)$$

Como realizado para as regiões interna e externa, a solução em primeira ordem também nos fornecerá a expressão para o campo elétrico na região 2:

$$E_2(z) = -\frac{d\phi_2(z)}{dz} = -\sqrt{2G\phi_2(z) + 2\frac{a_2}{b_2}\cosh[b_2\phi_2(z)] + \lambda_2} \quad (44)$$

Considerando que em $z \to h_g$, onde $h_g$ é a posição de fronteira entre o meio externo e o glicocálix, teremos então que o campo elétrico e o potencial nesse ponto serão:

$$E_i(z \to h_g) = E_g$$
$$\phi_i(z \to h_g) = \phi_g$$

logo:

$$\lambda_2 = -2\frac{a_i}{b_i}\cosh[b_i\phi_g] + E_g^2 \quad (45)$$

A partir da eq.(44) podemos obter a expressão para o potencial, mas tal qual a solução nas fases interna e externa, é necessário expandir o lado direito da eq(43) em série de Taylor, mas agora a análise se faz em torno de $\phi = \phi_g$, obtendo assim a relação:

$$\sqrt{2G\phi_2 + 2\frac{a_2}{b_2}\cosh[b_2\phi_2] + \lambda_2} = k_0 + k_1(\phi_2 - \phi_g) + k_2(\phi_2 - \phi_g)^2 + \ldots \quad (46)$$

onde as contantes dependem dos termos $G$, $a_2$, $b_2$ e $\phi_g$. Usando essa relação (46) em (34), chegamos a:

$$E_2(z) = -\frac{d\phi_2(z)}{dz} = k_0 + k_1(\phi_2(z) - \phi_g) + k_2(\phi_2(z) - \phi_g)^2 + \ldots \quad (47)$$

Como novamente os termos acima de segunda ordem pouco interferem em uma análise numérica, teremos:

$$\frac{d\phi_2(z)}{dz} = k_0 + k_1(\phi_2(z) - \phi_g) + k_2(\phi_2(z) - \phi_g)^2 \quad (48)$$

Resolvendo a equação (48), obtemos a expressão para o potencial elétrico dentro da região do glicocálix:

$$\phi_2(z) = \phi_{g1} + \frac{\sqrt{4k_0k_2 - k_1^2}}{2k_2}\tan\left(\sqrt{4k_0k_2 - k_1^2}\frac{(z - h_g)}{2} + \arctan\left(\frac{k_1}{\sqrt{4k_0k_2 - k_1^2}}\right)\right) - \frac{k_1}{2k_2} \quad (49)$$

onde $\phi_{g1}$ é o potencial na superfície que separa o meio intracelular e o glicocálix.

### 6. Solução para a bicamada

Como consideramos que a equação de Poisson para a região da bicamada é dada pela expressão (13), teremos então na direção $z$:

$$\frac{d^2\phi_2(z)}{dz^2} = 0 \quad (50)$$

A solução dessa equação nos fornece a expressão para o campo elétrico nessa região:

$$E_3(z) = -\frac{d\phi_2(z)}{dz} = -l_0 \quad (51)$$

e para o potencial obtemos:

$$\phi_3(z) = l_0 z + l_1 \quad (52)$$

onde as constantes $l_0$ e $l_1$ serão determinadas pelas condições de contorno apresentadas pela equação (15) e (16). Como nas regiões que limitam a bicamada os campos elétricos e os respectivos potenciais devem ser $E_1$, $\phi_1 = \phi_{g2}$, em $z \to -h/2$ e refere-se ao campo e o potencial na superfície que separa o glicocálix da bicamada, $E_2$, $\phi_2$, em $z \to h/2$, respectivamente, e refere-se ao potencial e o campo na camada que separa a bicamada do meio externo. Sendo assim a eq.(52) é reescrita como:

$$\phi_3(z) = -\frac{(E_{g1} + E_2)}{2}z + \frac{(\phi_2 + \phi_{g1})}{2} \qquad (53)$$

## 7. Discussão e Conclusões

As equações obtidas para o meio interno e externo da membrana do neurônio, apesar de meios com as mesmas características, contendo eletrólitos livres, possuem diferenças marcantes em sua expressões. Isso deve-se ao valores característicos do potencial em cada lado, não sendo inconsistente com o observado experimentalmente. A equação (28) que descreve o potencial no meio externo sugere a queda exponencial do mesmo, enquanto para o meio interno, equação (33), ocorre um pequena subida, de forma quadrática, a medida que ocorre uma entrada na membrana.

Para a região do glicocálix e da bicamanda foram encontradas expressões que dentro do limite esperado possuem perfil retilíneo, indicando que deve existir uma queda do potencial dentro dessas regiões de forma linear. A queda esperada para esse potencial, deve-se ao fato do valor convencionado para o potencial na região interna a membrana, que é de -70 mV.

## 8. Referências Bibliográficas